\documentclass[twocolumn,prx,aps,amsmath,amssymb,longbibliography]{revtex4-1}

\usepackage{graphicx}
\usepackage{dcolumn}
\usepackage{bm}

\begin{document}

\title{Dynamic Majorana resonances and universal symmetry of nonequilibrium
  thermoelectric quantum noise}

\author{Sergey Smirnov}
\affiliation{P. N. Lebedev Physical Institute of the Russian Academy of
  Sciences, 119991 Moscow, Russia}
\email{1) sergej.physik@gmail.com\\2)
  sergey.smirnov@physik.uni-regensburg.de\\3) ssmirnov@sci.lebedev.ru}

\date{\today}

\begin{abstract}
Nonequilibrium states of a nanoscopic system may be achieved by both applying
a bias voltage $V$ to its contacts and producing a difference $\Delta T$ in
their temperatures. Then the total current results from two competing flows,
induced by $V$ and $\Delta T$, respectively. Here we explore finite frequency
quantum noise of this thermoelectric current flowing through a quantum dot
whose low-energy dynamics is governed by Majorana degrees of freedom. We
demonstrate that at finite frequency $\omega$ Majorana zero modes induce a
perfect universal symmetry between photon emission and absorption spectra and
produce universal thermoelectric resonances in their frequency dependence in
contrast to non-Majorana quantum noise which is either asymmetric or
non-universal. In particular, at low temperatures the differential
thermoelectric quantum noise induced by Majorana zero modes shows resonances
with a nontrivial maximum $\frac{e^3}{h}\log(2^{1/4})$ at
$\omega=\mp\frac{|eV|}{\hbar}$ when $k_\text{B}\Delta T\ll|eV|$. Our results
challenge cutting-edge experiments using quantum noise detectors to reveal the
universal spectral symmetry and resonant structure of the Majorana
thermoelectric finite frequency quantum noise.
\end{abstract}

\maketitle

\section{Introduction}\label{intro}
Quantum materials possessing various topological superconducting phases admit,
among others, specific phases characterized by the presence of Majorana zero
modes (MZMs). In such phases, achieved via quantum phase transitions, some pairs of
Majorana fermions break during a quantum phase transition and form MZMs
localized at interfaces separating topologically different systems. While in
the particle physics Majorana fermions \cite{Majorana_1937} are Abelian, their
condensed matter implementation assumes non-Abelian nature of MZMs which are
highly nonlocal quasiparticles. In particular, MZMs appear at the opposite
ends of one-dimensional systems effectively implementing
\cite{Alicea_2012,Flensberg_2012,Sato_2016,Aguado_2017,Lutchyn_2018} the
topological superconducting phase of the Kitaev model \cite{Kitaev_2001}
using topological insulators \cite{Fu_2008,Fu_2009} or spin-orbit coupled
semiconductors \cite{Lutchyn_2010,Oreg_2010}.

Experiments \cite{Mourik_2012,Albrecht_2016,Zhang_2018} on detections of
Majorana signatures usually analyze quantum transport through a mesoscopic
system, more specifically the mean electric current providing the differential
conductance. On one side, such transport experiments, although of high
quality, might be opposed to experiments based on braiding MZMs for fully
conclusive claims whether MZMs are present in a given system. On the other
side, recent success in thermodynamic and transport experiments
\cite{Hartman_2018,Kleeorin_2019} measuring the entropy of a mesoscopic system
is very appealing for future experimental detection (see, {\it e.g.}, Ref.
\cite{Sela_2019}) of Majorana tunneling entropy \cite{Smirnov_2015} to avoid
any braiding and at the same time to unambiguously reveal MZMs in mesoscopic
setups.

Braidings are, however, necessary for practical applications of MZMs in
devices implementing topological quantum computations \cite{Kitaev_2003}. The
final result of a sequence of braid operations depends only on its
topology. It is thus robust against many perturbations which have a local
character and in this way implements fault-tolerant quantum computation
protected from major sources of decoherence via global topologically
nontrivial properties \cite{Nayak_2008}. Remaining sources of decoherence
include noise which is unavoidable in practical realizations of Majorana
braiding influenced by an external environment of a realistic device
\cite{Pedrocchi_2015}.

In turn, in mesoscopic setups MZMs determine not only the mean values but also
lead to significant noise of physical observables, such as the electric
current, both in the static limit
\cite{Liu_2015,Liu_2015a,Beenakker_2015,Haim_2015,Smirnov_2017}, and at finite
frequencies \cite{Valentini_2016,Bathellier_2019,Smirnov_2019}. Measuring
noise of nonequilibrium electric currents provides at finite frequencies
$\omega$ photon emission ($\omega<0$) and absorption ($\omega>0$)
spectra. These spectra, as shown below, are either asymmetric, when the
Majoranas strongly overlap and form partially separated Andreev bound states,
or symmetric but not universal, when the Majoranas are absent at all as in
trivial quantum dot (QD) systems. The degree of this {\it spectral asymmetry}
or {\it non-universality} is relatively easy to verify in contemporary
experiments via dependence of the spectra on the gate voltage controlling the
chemical potential.
\begin{figure*}
\includegraphics[width=16.0 cm]{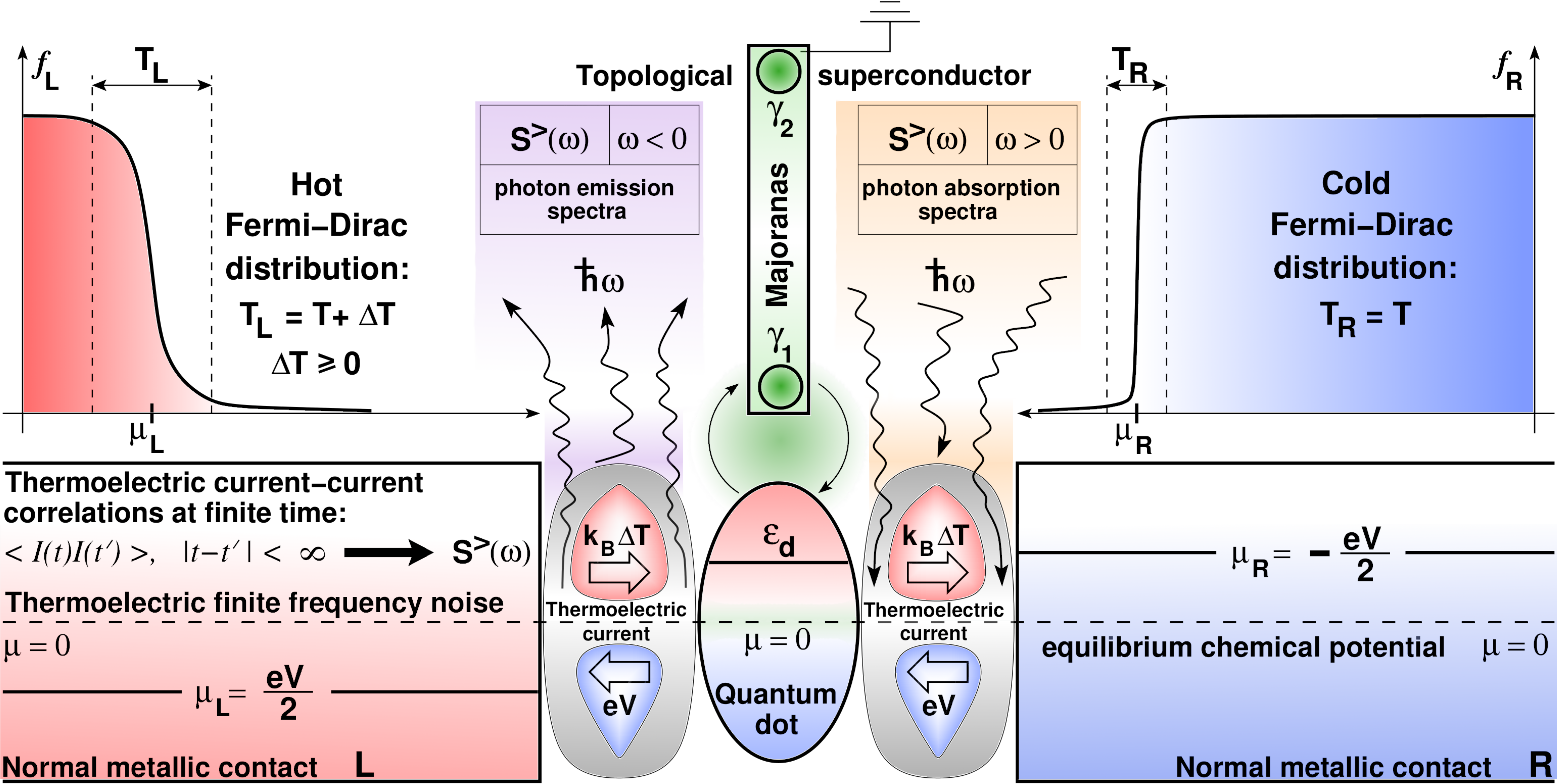}
\caption{\label{figure_1} A sketch of a realistic QD device where both a bias
  voltage $V$ and a temperature difference $\Delta T$ are the sources of
  nonequilibrium states of a QD. The QD is coupled via tunneling to two
  massive normal metals, the left ($L$) and right ($R$) contacts. The gray
  areas between the QD and contacts specify the nonequilibrium thermoelectric
  current incorporating two competing flows (big right-to-left and
  left-to-right arrows), excited, respectively, by $V$ (the blue droplet) and
  by $\Delta T$ (the red droplet). The device includes a topological
  superconductor (TS) with two MZMs $\gamma_{1,2}$ at its ends (two green
  circles). The quasiparticle flows, denoted by the two thin vertical arc
  arrows around the green area, entangle the QD states with the Majoranas via
  tunneling processes involving $\gamma_1$.}
\end{figure*}

In this paper we demonstrate that the above picture, which would commonly be
observed for partially separated Andreev bound states or trivial systems,
changes drastically when MZMs are involved in the competition between two
nonequilibrium quasiparticle flows, excited, respectively, by a bias voltage
$V$ and a temperature difference $\Delta T$. Remarkably, the quasiparticle
fluctuations, resulting in finite frequency quantum noise of the induced
thermoelectric current, are shown to acquire {\it universal spectral symmetry}
and {\it nontrivial resonant structure}.

Specifically, we present results for thermoelectric transport through a QD
whose states are entangled with a Kitaev's chain in its topological
superconducting phase supporting two MZMs at its ends as shown in
Fig. \ref{figure_1}. The QD also interacts with two normal metals, left and
right contacts, whose chemical potentials and temperatures are denoted,
respectively, as $\mu_{L,R}$ and $T_{L,R}$, $\mu_L=-\mu_R=eV/2$
($eV<0$), $T_L=T+\Delta T$ ($\Delta T\geqslant 0$), $T_R=T$, {\it i.e.} the
left (right) contact is hot (cold). Instead of the temperature difference it
is convenient to use the thermal voltage $eV_T\equiv k_\text{B}\Delta T$.

While the static ($\omega=0$) limit \cite{Smirnov_2018} of Majorana
thermoelectric noise provides fluctuation physics at long times, it does not
have access to the corresponding photon emission and absorption spectra which
are of direct experimental interest in the quantum limit \cite{Clerk_2010}. In
contrast, at finite frequencies quantum fluctuations $\delta I(t)$ of the
thermoelectric current $I(t)$ keep inside wealth of information about the
photon spectra never explored in thermoelectric Majorana setups.

Here, for the case of the setup in Fig. \ref{figure_1}, we explore the
properties of these Majorana finite frequency thermoelectric quantum
fluctuations encoded in the photon absorption and emission spectra obtained
via the Fourier transform, $S^>(\omega,V,V_T)$, of the greater noise
correlation function in the left contact,
$S^>(t-t',V,V_T)=\langle\delta I_L(t)\delta I_L(t')\rangle$. To focus on
universal properties we compute the differential thermoelectric finite
frequency quantum noise, $\partial S^>(\omega,V,V_T)/\partial V_T$, which has
universal units of $e^3/h$. In particular, when the MZMs are involved in the
low temperature interplay between the two quasiparticle flows induced by,
respectively, $V$ and $V_T$, we predict (1) that the differential
thermoelectric quantum noise (DTQN) acquires universal symmetry,
$\partial S^>(-\omega,V,V_T)/\partial V_T=\partial S^>(\omega,V,V_T)/\partial V_T$;
(2) in the regime $eV_T\gg|eV|$ it has a resonance located around $\omega=0$
with a nontrivial universal maximum $[1+\log(2)](e^3/h)$ reached exactly at
$\omega=0$; (3) in the regime $eV_T\ll|eV|$ the resonance located around
$\omega=0$ persists but its maximum lowers down to another nontrivial
universal maximum $[1+\log(2^{1/2})](e^3/h)$ also reached exactly at
$\omega=0$; (4) in the same regime DTQN exhibits resonances located around
$\omega=\mp|eV|/\hbar$ with one more nontrivial universal maximum
$[\log(2^{1/4})](e^3/h)$ reached exactly at $\omega=\mp|eV|/\hbar$.

The paper is organized as follows. In Section \ref{tmtqn} we discuss in detail
the theoretical model involving MZMs and define the specific observable we
explore, that is the thermoelectric quantum noise. The results are presented
and thoroughly analyzed in Section \ref{rmtqn} where the emphasis is made on
universal behavior of the thermoelectric quantum noise. With Section
\ref{concl} we provide realistic parameters in the SI units and conclude that
the universal results from Section \ref{rmtqn} are well achievable in modern
labs.
\section{Theoretical model and thermoelectric quantum noise}\label{tmtqn}
To explore the essential aspects of quantum thermoelectric transport we
describe the setup shown in Fig. \ref{figure_1} within a theoretical model
specified by the Hamiltonians of the QD, contacts, TS, the Hamiltonians for
the tunneling between the QD and contacts and for the tunneling between the QD
and TS.

The QD has one nondegenerate single-particle level $\epsilon_d$. A gate
voltage may tune the position of $\epsilon_d$ with reference to the chemical
potential $\mu$. The corresponding noninteracting Hamiltonian is
$\hat{H}_d=\epsilon_d d^\dagger d$.

The normal metallic contacts are assumed to be noninteracting and large enough
so that their energy spectra are continuous,
$\hat{H}_c=\sum_{l=L,R}\sum_k\epsilon_k c^\dagger_{lk}c_{lk}$. Within energy
ranges relevant for transport one traditionally assumes a constant density of
states $\nu_c/2$. The contacts are in equilibrium with the corresponding
Fermi-Dirac distributions,
$f_{L,R}(\epsilon)=\{\exp[(\epsilon-\mu_{L,R})/T_{L,R}]+1\}^{-1}$.

At low energies one may effectively capture relevant physics of the TS via the
Hamiltonian $\hat{H}_{tsc}=i\xi\gamma_2\gamma_1/2$ represented in terms of the
Majorana operators $\gamma_{1,2}$ satisfying the conjugation relation
$\gamma_{1,2}^\dagger=\gamma_{1,2}$ and the anticommutation relations of the
Clifford algebra, $\{\gamma_i,\gamma_j\}=2\delta_{ij}$. The energy $\xi$
characterizes the degree of the overlap of the MZMs. Small values of $\xi$
correspond to situations when the two Majoranas are well separated while for
large values of $\xi$ the two MZMs merge into a single Dirac fermion.
\begin{figure}
\includegraphics[width=8.0 cm]{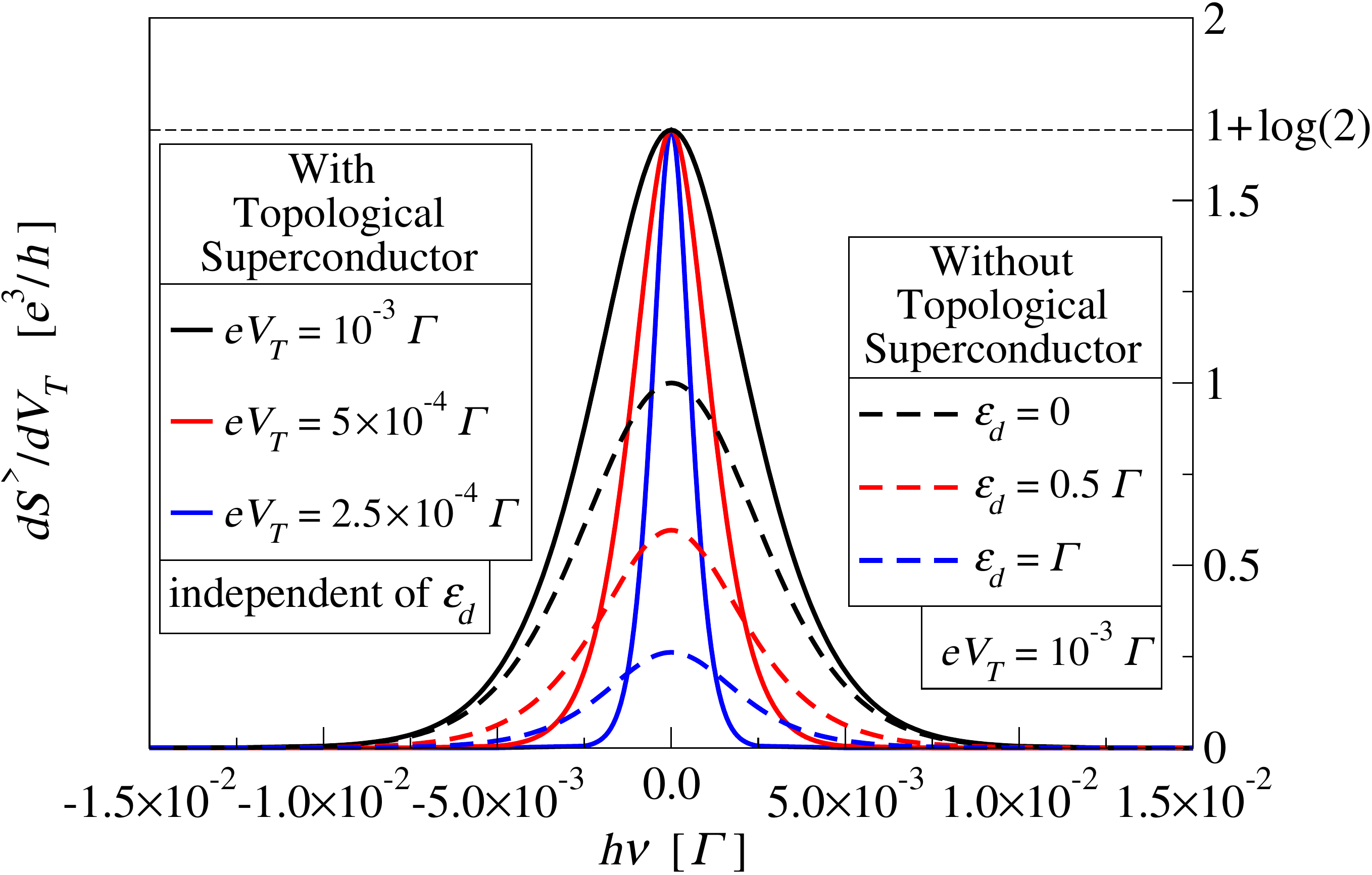}
\caption{\label{figure_2} DTQN $\partial S^>(\nu,V,V_T)/\partial V_T$ as a
  function of the frequency $\nu$ ($h\nu=\hbar\omega$) is shown in the regime
  $eV_T\gg|eV|$. Here $k_\text{B}T/\Gamma=10^{-10}$, $|\eta|/\Gamma=10^3$,
  $\xi/\Gamma=10^{-4}$ and $|eV|/\Gamma=10^{-5}$. The three solid curves
  demonstrate a resonance of $\partial S^>(\nu,V,V_T)/\partial V_T$ located
  around $\nu=0$ for three different values of $eV_T$. The three dashed curves
  show the spectra of the trivial QD system, $\xi=0$, $\eta=0$, for three
  different values of the gate voltage.}
\end{figure}

The tunneling between the QD and contacts as well as between the QD and TS may
be taken into account via, respectively, the Hamiltonians
$\hat{H}_{d-c}=\sum_{l=L,R}\sum_k(\mathcal{T}c^\dagger_{lk}d+\mathcal{T}^*d^\dagger c_{lk})$
and $\hat{H}_{d-tsc}=\gamma_1(\eta d-\eta^*d^\dagger)$. The intensities of
the tunneling mechanisms $\hat{H}_{d-c}$ and $\hat{H}_{d-tsc}$ are
characterized by the energies $\Gamma=2\pi\nu_c|\mathcal{T}|^2$ and $|\eta|$,
respectively.

The quantum noise of the nonequilibrium thermoelectric current may
conveniently be obtained using the Keldysh generating functional
\cite{Altland_2010},
\begin{equation}
\begin{split}
&Z[J_l(t)]=\int\mathcal{D}[\bar{\theta}(t),\theta(t)]e^{\frac{i}{\hbar}S_K[\bar{\theta}(t),\theta(t);J_l(t)]},\\
&\{\theta(t)\}=\{\psi(t),\phi_{lk}(t),\zeta(t)\},
\end{split}
\label{Keld_gen_func}
\end{equation}
with the Grassmann fields of the QD, $\psi(t)$, contacts, $\phi_{lk}(t)$, and
TS, $\zeta(t)$, defined on the Keldysh closed time contour,
$t\in\mathcal{C}_K$. In particular, to analyze correlations between the values
$I_l(t)$ and $I_{l'}(t')$ of the thermoelectric current at different instants
of time, $t$ and $t'$, one may introduce the source fields $J_{lq}(t)$
(subscript $q$ denotes the forward/backward branch of $\mathcal{C}_K$,
$q=\pm$),
\begin{equation}
\begin{split}
&S_{scr}=-\int_{-\infty}^\infty dt\sum_{l=L,R}\sum_{q=+,-}J_{lq}(t)I_{lq}(t),\\
&I_{lq}(t)=\frac{ie}{\hbar}\sum_k[\mathcal{T}\bar{\phi}_{lkq}(t)\psi_q(t)-\mathcal{T}^*\bar{\psi}_q(t)\phi_{lkq}(t)],
\end{split}
\label{Scr_act}
\end{equation}
\begin{equation}
\begin{split}
&\langle I_l(t)I_{l'}(t')\rangle=(i\hbar)^2\frac{\delta^2Z[J_l(t)]}{\delta J_{l-}(t)\delta J_{l'+}(t')}\biggl|_{J_{lq}(t)=0}.
\end{split}
\label{Curr_curr_corr}
\end{equation}

We focus on the greater noise correlation function in the left contact,
$S^>(t-t';V,V_T)=\langle I_L(t)I_L(t')\rangle-I^2_L(V,V_T)$, where
$I_L(V,V_T)$ is the mean thermoelectric current. Due to the stationary
conditions it depends only on the difference, $(t-t')$, of the instants of
time and thus its Fourier transform depends on only one frequency $\omega$ (or
$\nu$, $h\nu=\hbar\omega$). One then obtains the photon emission/absorption
spectra,
\begin{equation}
S^{em/ab}(\nu,V,V_T)=
\begin{cases}
S^>(\nu,V,V_T),\quad\nu<0\\
S^>(\nu,V,V_T),\quad\nu>0
\end{cases},
\label{Ab_em_spec}
\end{equation}
characterizing the thermoelectric transport.
\begin{figure}
\includegraphics[width=8.0 cm]{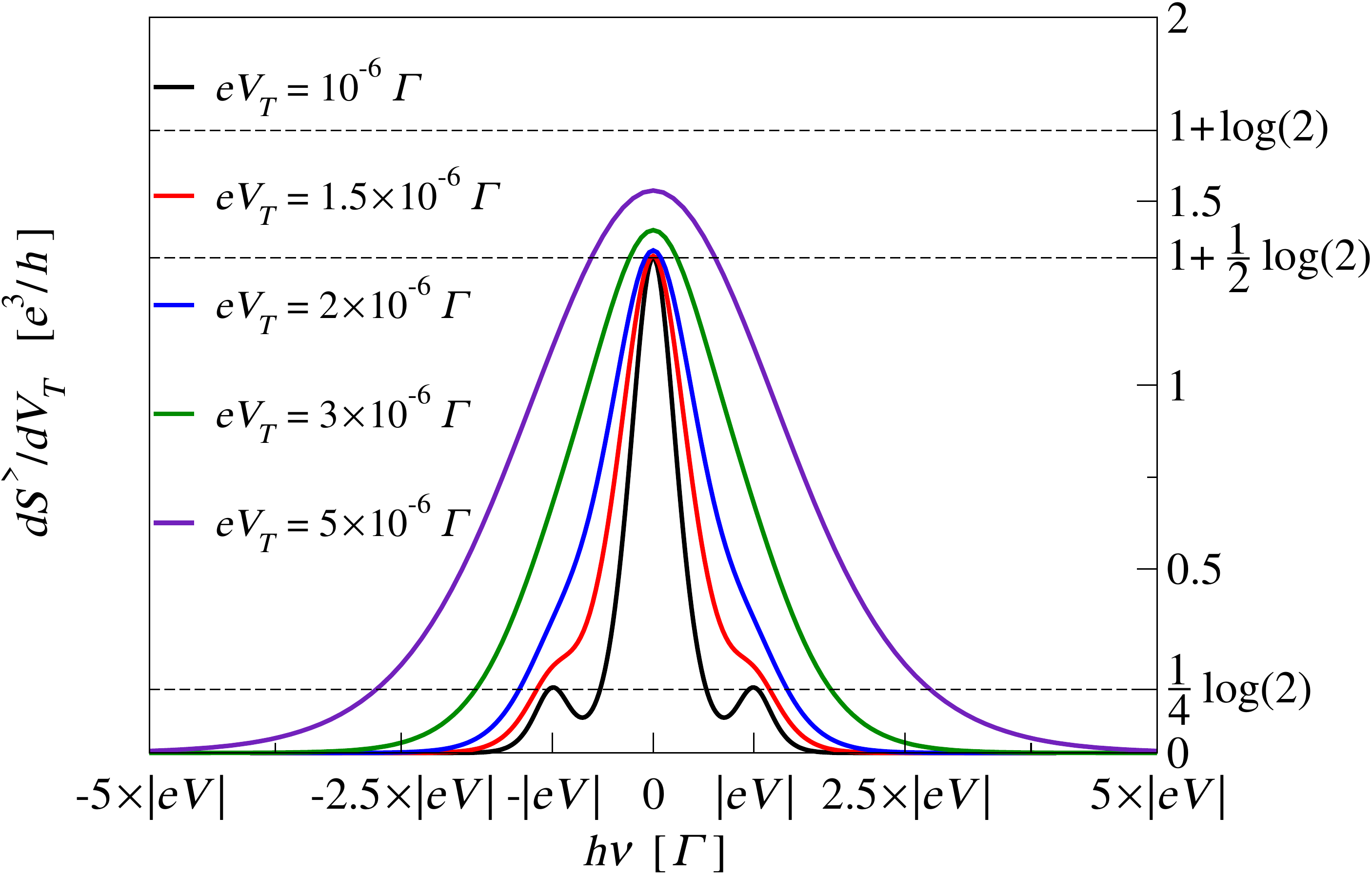}
\caption{\label{figure_3} DTQN $\partial S^>(\nu,V,V_T)/\partial V_T$ as a
  function of the frequency $\nu$ is shown in the regimes $eV_T<|eV|$:
  $eV_T=0.5|eV|$ (purple), $eV_T=0.3|eV|$ (green), $eV_T=0.2|eV|$ (blue) and
  $eV_T\ll|eV|$: $eV_T=0.15|eV|$ (red), $eV_T=0.1|eV|$ (black). The other
  parameters have the same values as in Fig. \ref{figure_2}.}
\end{figure}

Below our results reveal universality of Majorana thermoelectric quantum
fluctuations at finite frequencies via numerical computations of the
DTQN having universal units of $e^3/h$. We assume the strong Majorana
tunneling regime characterized by large values of $|\eta|$, that is
\begin{equation}
|\eta|\geqslant\text{max}\{|\epsilon_d|,|eV|,eV_T,k_\text{B}T,\xi\}.
\label{M_reg}
\end{equation}

Finally, to avoid any misconception, it is important to keep in mind that we
explore the quantum noise (see also Refs. \cite{Lesovik_1997,Gavish_2000}).
One should not mix it up with the classical shot noise having different
physical nature. The latter is given as the symmetrized current-current
correlator. As a result, the classical shot noise has symmetric spectra and
any discussion of its spectral asymmetry would obviously make no sense.
\section{Results for the Majorana thermoelectric quantum noise}\label{rmtqn}
First, we show results in the regime $eV_T\gg|eV|$. In this regime there
appears a resonance around $\nu=0$ shown by the solid curves in
Fig. \ref{figure_2}. The maximum of this resonance is reached at $\nu=0$ and
takes the universal nontrivial value
$\partial S^>(\nu=0,V,V_T)/\partial V_T=[1+\log(2)](e^3/h)$. In the regime
$|eV|\ll eV_T$ this maximum does not depend on $\epsilon_d$, $V$ and
$V_T$. The full width of this resonance at half of its maximum is proportional
to $eV_T$. Moreover, the resonance demonstrates the full symmetry between the
photon emission ($\nu<0$) and photon absorption ($\nu>0$) spectra, {\it i.e.}
$\partial S^>(-\nu,V,V_T)/\partial V_T=\partial S^>(\nu,V,V_T)/\partial V_T$.
In contrast, for the trivial QD system having no TS, $\xi=0$, $\eta=0$, the
spectra, shown by the dashed curves, are not universal since they depend on
$\epsilon_d$.

When $eV_T$ becomes comparable and below $|eV|$, the
resonance shown in Fig. \ref{figure_2} starts to change. This is shown in
Fig. \ref{figure_3} for various values of $eV_T$, from $eV_T<|eV|$ down to
$eV_T\ll|eV|$. The resonance persists in all the curves having the spectral
symmetry,
$\partial S^>(-\nu,V,V_T)/\partial V_T=\partial S^>(\nu,V,V_T)/\partial V_T$.
The maximum of the resonance is at $\nu=0$ for all the curves. For $eV_T<|eV|$
the maximum is not universal: it decreases from the universal nontrivial value
$[1+\log(2)](e^3/h)$, reached in the regime $eV_T\gg|eV|$ (see
Fig. \ref{figure_2}), down to the universal nontrivial value
$[1+\log(2^{1/2})](e^3/h)$, reached in the regime $eV_T\ll|eV|$. Moreover,
when $eV_T$ becomes much below $|eV|$, in the vicinity of $h\nu=\mp|eV|$ there
\begin{figure}
\includegraphics[width=8.0 cm]{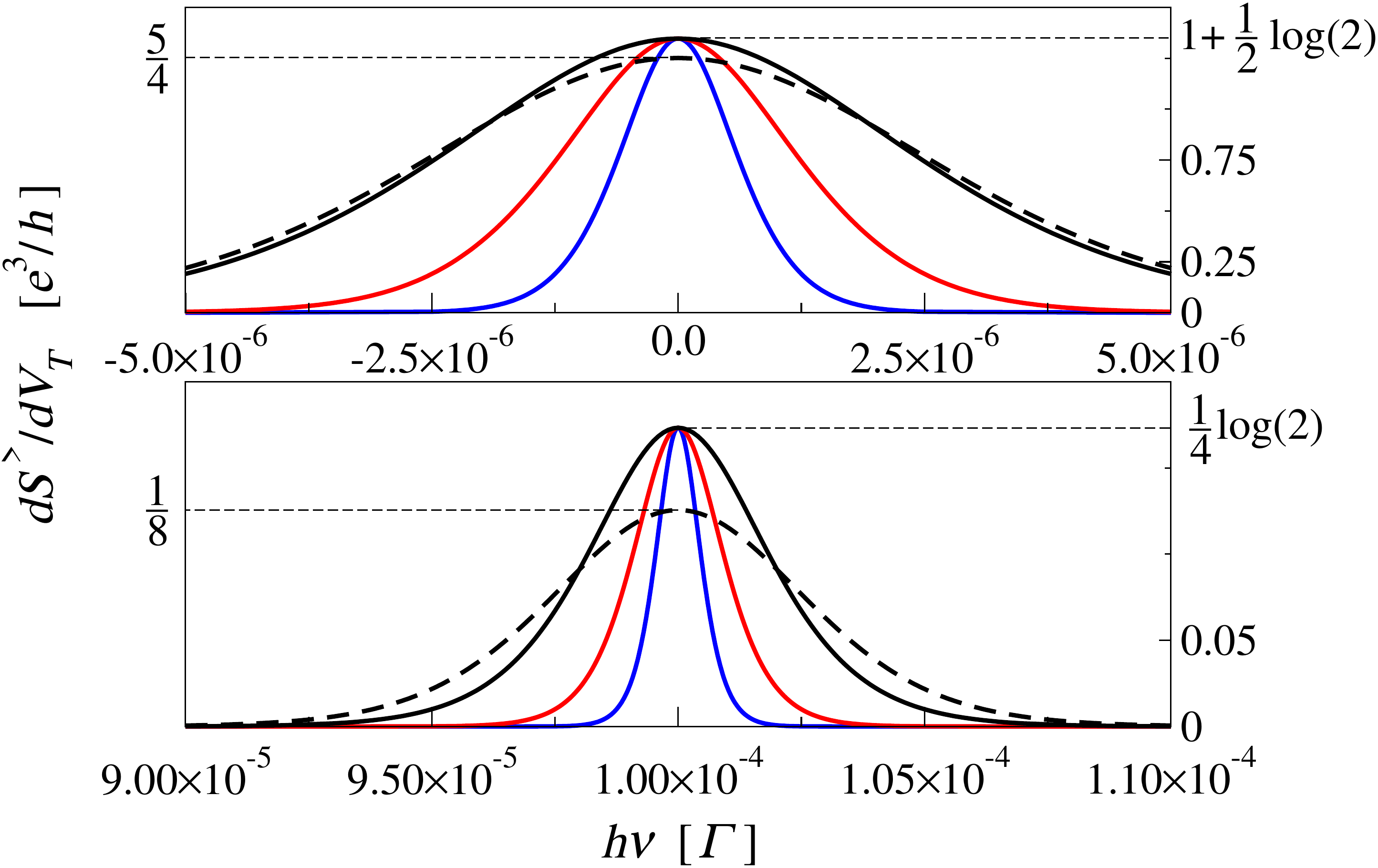}
\caption{\label{figure_4} The resonances of DTQN,
  $\partial S^>(\nu,V,V_T)/\partial V_T$, located around $h\nu=0$ (upper
  panel) and $h\nu=|eV|$ (lower panel), are shown as functions of the
  frequency $\nu$ when $|eV|$ essentially exceeds $eV_T$. Here
  $|eV|/\Gamma=10^{-4}$ while $eV_T=10^{-2}|eV|$ (black curves),
  $eV_T=5\cdot 10^{-3}|eV|$ (red curves) and $eV_T=2.5\cdot 10^{-3} |eV|$
  (blue curves). The dashed curves show the universal results for higher
  temperatures, in the regime $eV_T\ll k_\text{B}T\ll|eV|$, with
  $eV_T=10^{-4}|eV|$, $k_\text{B}T=10^{-2}|eV|$. The other parameters have the
  same values as in Fig. \ref{figure_2}.}
\end{figure}
\begin{figure}
\includegraphics[width=8.0 cm]{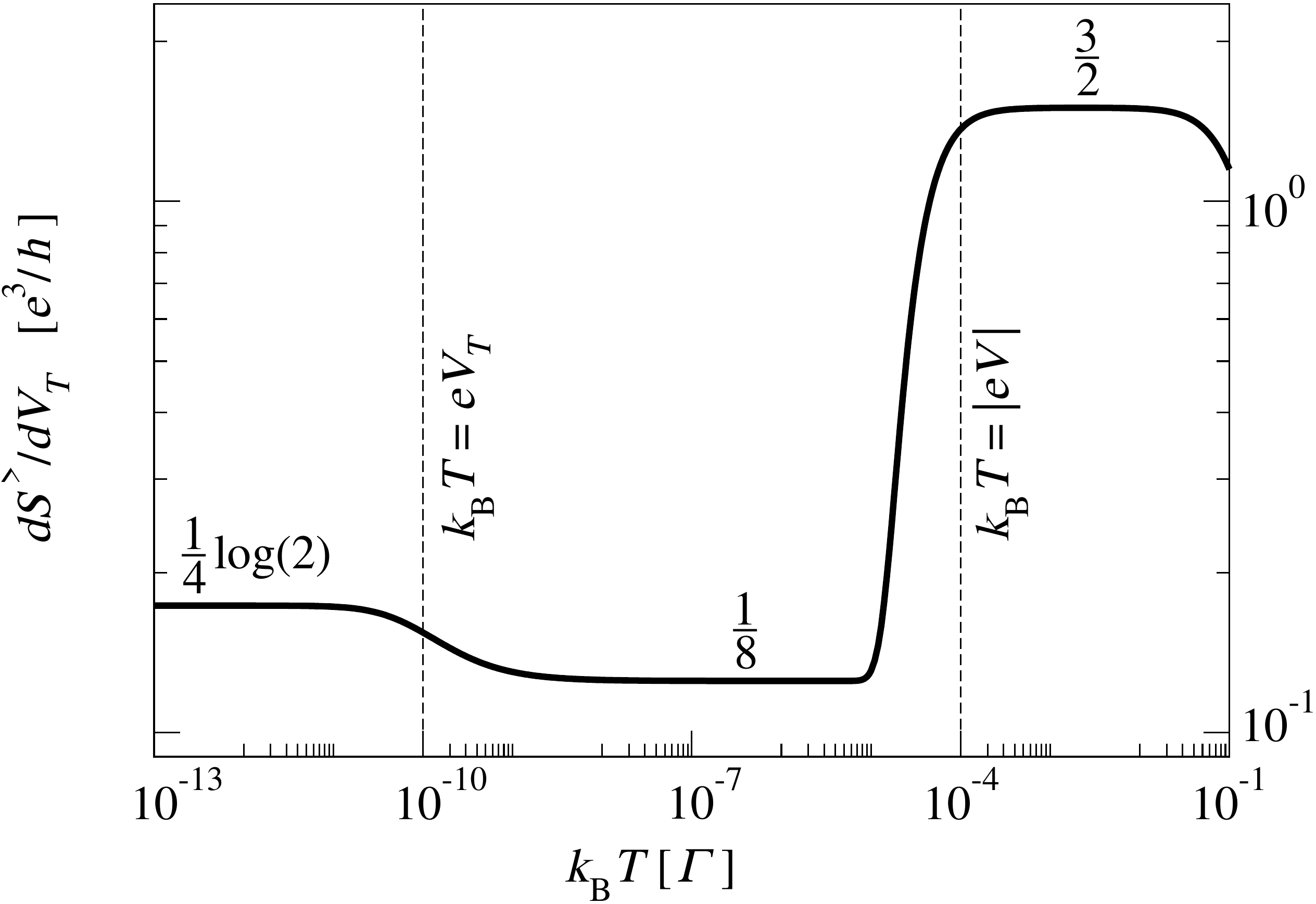}
\caption{\label{figure_5} The universal temperature dependence of DTQN,
  $\partial S^>(\nu,V,V_T)/\partial V_T$, at $h\nu=|eV|$. Here
  $|eV|/\Gamma=10^{-4}$ while $eV_T=10^{-6}|eV|$. The other parameters have
  the same values as in Fig. \ref{figure_2}.}
\end{figure}
develop two shoulders (red curve, $eV_T=0.15|eV|$). If $eV_T$ is decreased
still further, the shoulders turn into nonequilibrium resonances located
around $h\nu=\mp|eV|$ (black curve, $eV_T=0.1|eV|$). For $eV_T\ll|eV|$ the
maxima of these resonances are at $h\nu=\mp|eV|$ and take the universal
nontrivial value $[\log(2^{1/4})](e^3/h)$. As shown in Fig. \ref{figure_4} by
solid curves, the full widths of the resonances, located around $h\nu=0$ and
$h\nu=|eV|$, at half of their maxima are proportional to $eV_T$. The
universality of the maxima $[1+\log(2^{1/2})](e^3/h)$ and
$[\log(2^{1/4})](e^3/h)$, located, respectively, at exactly $h\nu=0$ (upper
panel) and $h\nu=|eV|$ (lower panel), manifests in their independence of
$\epsilon_d$, $V$ and $V_T$. As the temperature is increased, the universal
spectral symmetry persists and one crosses over through two new regimes,
$eV_T\ll k_\text{B}T\ll|eV|$ and $eV_T\ll|eV|\ll k_\text{B}T$. When
$eV_T\ll k_\text{B}T\ll|eV|$, the resonances located around $h\nu=0$ and
$h\nu=\mp|eV|$ persist but, as it is shown by the dashed lines in
Fig. \ref{figure_4}, their maxima acquire new universal values, $5e^3/4h$ and
$e^3/8h$, respectively, and the full widths of the resonances at half of their
maxima are proportional to $k_\text{B}T$. If the temperature
\begin{figure}
\includegraphics[width=8.0 cm]{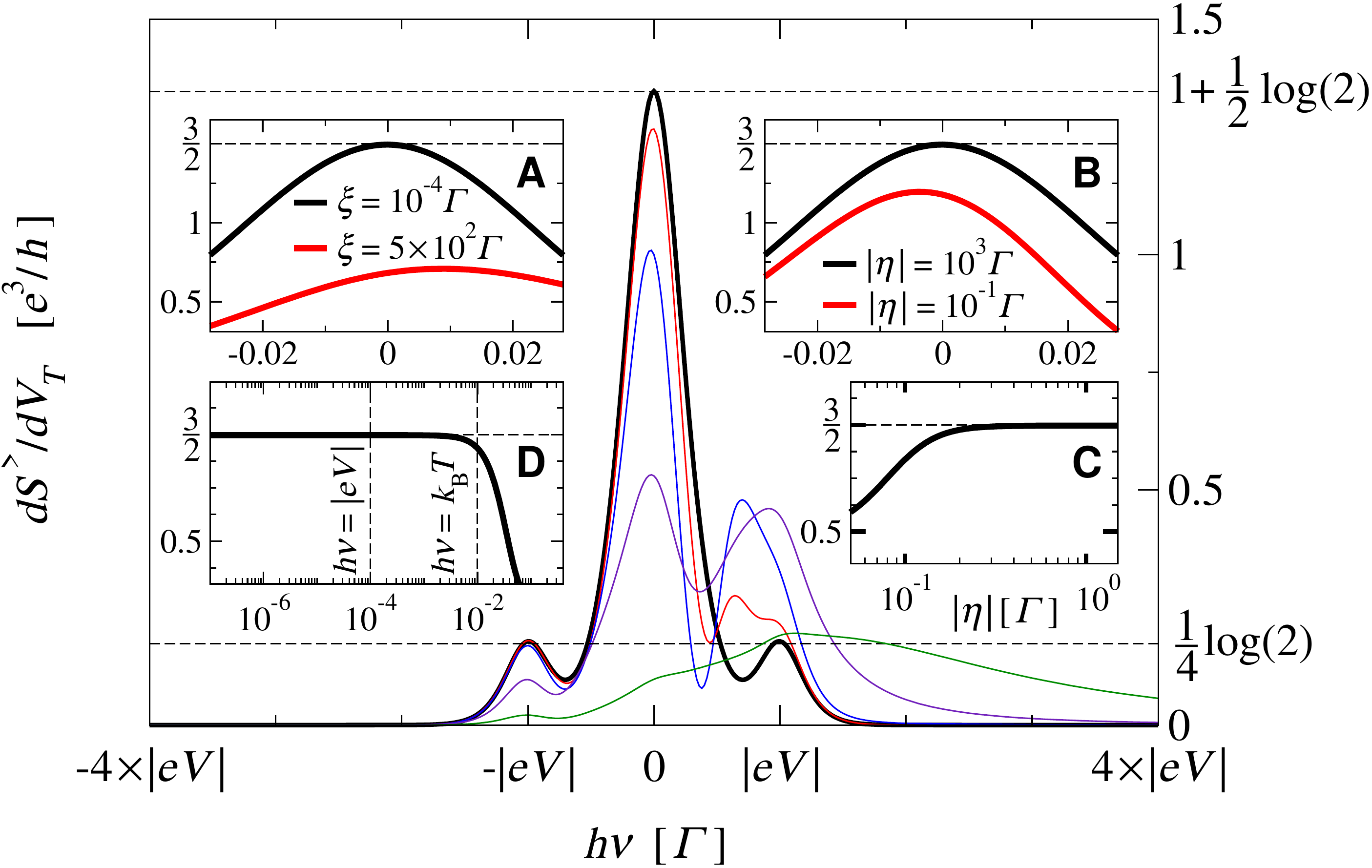}
\caption{\label{figure_6} DTQN
  $\partial S^>(\nu,V,V_T)/\partial V_T$ as a function of the frequency $\nu$
  is shown for $eV_T=0.1|eV|$. Here $\xi/\Gamma=10^{-4}$ (black),
  $\xi/\Gamma=0.1$ (red), $\xi/\Gamma=0.2$ (blue), $\xi/\Gamma=0.4$ (purple),
  $\xi/\Gamma=0.8$ (green). The other parameters have the same values as in
  Fig. \ref{figure_2}. Inset A: $\partial S^>(\nu,V,V_T)/\partial V_T$ for two
  values of $\xi$ in the regime $eV_T\ll|eV|\ll k_\text{B}T$ with
  $eV_T=10^{-6}|eV|$, $k_\text{B}T=10^2|eV|$, $|eV|/\Gamma=10^{-4}$ and
  $|\eta|/\Gamma=10^3$. Inset B: the same as in Inset A but for two values of
  $|\eta|$ and $\xi/\Gamma=10^{-4}$. Inset C:
  $\partial S^>(\nu,V,V_T)/\partial V_T$ at $h\nu=|eV|$ as a function of
  $|\eta|$ with other parameters as in Inset B. Inset D: the same as in
  Inset A but with $|\eta|=\Gamma$, $\xi/\Gamma=10^{-4}$.}
\end{figure}
is increased still further, the resonances at $h\nu=0$ and $h\nu=\mp|eV|$
widen and finally merge into a single resonance located around $h\nu=0$. The
full width of this resonance at half of its maximum is proportional to
$k_\text{B}T$. When $eV_T\ll|eV|\ll k_\text{B}T$, the maximum of this
resonance is exactly at $h\nu=0$ and takes the universal value $3e^3/2h$. In
this regime DTQN at $h\nu=\mp|eV|$ with very high precision is equal to its
value at $h\nu=0$, {\it i.e.} to $3e^3/2h$. DTQN at $h\nu=|eV|$ is shown in
Fig. \ref{figure_5} as a universal function of the temperature over the three
regimes $k_\text{B}T\ll eV_T\ll|eV|$, $eV_T\ll k_\text{B}T\ll|eV|$ and
$eV_T\ll|eV|\ll k_\text{B}T$.

Let us look at what happens with the above universal spectral symmetry and
resonant structure when MZMs significantly overlap or when the coupling
strength to the TS is varied. The overlap of MZMs is achieved via increasing
$\xi$. Note that our model with large values of $\xi$ may also be interpreted
as the one where the QD is coupled to a partially separated Andreev bound
state \cite{Moore_2018} localized near the corresponding end of the TS (see
also Ref. \cite{Hell_2018}). We assume that the Majoranas, composing the
partially separated Andreev bound state, significantly overlap but are still
well separated in order to neglect the coupling of the second Majorana,
$\gamma_2$, to the QD. A more general situation with finite couplings of
$\gamma_1$ and $\gamma_2$ to the QD (see, {\it e.g.}, Ref. \cite{Clarke_2017})
will be explored in our future research. We also note that, although our model
relying upon energy spectra is quite realistic, in experiments it may become
necessary to introduce a measure for spatial separation of the two Majoranas
\cite{Deng_2018}.

Fig. \ref{figure_6} presents results for $eV_T=0.1|eV|$ when the triple
resonant structure is present for extremely well ($\xi/\Gamma=10^{-4}$)
separated MZMs. The thin curves illustrate how this perfectly symmetric
structure (thick black curve) gains asymmetry when $\xi$ increases and the two
Majoranas compose a partially separated Andreev bound state. Insets A-D show
results in the regime $eV_T\ll|eV|\ll k_\text{B}T$. In the insets we specify
$\epsilon_d=0.5\Gamma$. Inset A shows that also at high temperatures the
perfect Majorana spectral symmetry, demonstrated by the black curve, is broken
by partially separated Andreev bound states, modeled via large values of
$\xi$, as demonstrated by the red curve. The black curve in Inset B is a
universal (independent of $\epsilon_d$) result demonstrating the Majorana
spectral symmetry when Inequality (\ref{M_reg}) is satisfied. For the red
curve in Inset B Inequality (\ref{M_reg}) is not satisfied and this
non-universal (dependent on $\epsilon_d$) curve is asymmetric. The
universality (independence of $\epsilon_d$) persists down to
$|\eta|\approx 0.3\Gamma$ as demonstrated in Inset C where DTQN at $h\nu=|eV|$
is shown as a function of $|\eta|$. This function significantly deviates from
$3e^3/2h$ for $|\eta|<0.3\Gamma$. As to the frequency dependence, one may
observe the universal plateau $3e^3/2h$ at $h\nu\lesssim k_\text{B}T$, as
shown in Inset D.
\section{Conclusion}\label{concl}
In conclusion, advantages in experimental testing of our results are
obvious. First, it is convenience in observing the universal spectral symmetry
which does not require measuring any specific value: the spectral symmetry is
either present or not and one immediately distinguishes between Majoranas and
other states like partially separated Andreev bound states. Second, at high
temperatures, $eV_T\ll|eV|\ll k_\text{B}T$, one measures the universal value
$3e^3/2h$ at realistic parameters. Indeed, as shown above, the universality is
reached when $|\epsilon_d|\leqslant|\eta|$ and $|\eta|\gtrsim 0.3\Gamma$,
{\it e.g.} $|\eta|=\Gamma$, as may be tuned by the corresponding gate
voltages. The universal value $3e^3/2h$ may be observed at
$k_\text{B}T/\Gamma=10^{-2}$, as in Inset D of Fig. \ref{figure_6}. The
largest energy scale is then $\Gamma$. It must be below the induced
superconducting gap $\Delta$: $\Gamma\lesssim\Delta$ or
$k_\text{B}T\lesssim 10^{-2}\Delta$. Refs. \cite{Mourik_2012,Zhang_2018} give
$\Delta\approx 250$ or $200\,\mu\text{eV}$. Thus
$T\lesssim 2.5\,\mu\text{eV}/k_\text{B}\approx 0.03\,\text{K}=30\,\text{mK}$.
Temperatures below $30\,\text{mK}$ are well achievable in modern labs. Since
quantum noise detectors have already been proposed
\cite{Averin_2000,Clerk_2004,Mozyrsky_2004}, our results are of direct
interest for contemporary experiments detecting in the quantum limit the
universal spectral symmetry and resonant structure of DTQN in Majorana
mesoscopic setups.
\section*{Acknowledgments}
The author thanks Milena Grifoni, Andreas K. H{\"u}ttel and Wataru Izumida for
useful discussions.


\begin{thebibliography}{41}%
\makeatletter
\providecommand \@ifxundefined [1]{%
 \@ifx{#1\undefined}
}%
\providecommand \@ifnum [1]{%
 \ifnum #1\expandafter \@firstoftwo
 \else \expandafter \@secondoftwo
 \fi
}%
\providecommand \@ifx [1]{%
 \ifx #1\expandafter \@firstoftwo
 \else \expandafter \@secondoftwo
 \fi
}%
\providecommand \natexlab [1]{#1}%
\providecommand \enquote  [1]{``#1''}%
\providecommand \bibnamefont  [1]{#1}%
\providecommand \bibfnamefont [1]{#1}%
\providecommand \citenamefont [1]{#1}%
\providecommand \href@noop [0]{\@secondoftwo}%
\providecommand \href [0]{\begingroup \@sanitize@url \@href}%
\providecommand \@href[1]{\@@startlink{#1}\@@href}%
\providecommand \@@href[1]{\endgroup#1\@@endlink}%
\providecommand \@sanitize@url [0]{\catcode `\\12\catcode `\$12\catcode
  `\&12\catcode `\#12\catcode `\^12\catcode `\_12\catcode `\%12\relax}%
\providecommand \@@startlink[1]{}%
\providecommand \@@endlink[0]{}%
\providecommand \url  [0]{\begingroup\@sanitize@url \@url }%
\providecommand \@url [1]{\endgroup\@href {#1}{\urlprefix }}%
\providecommand \urlprefix  [0]{URL }%
\providecommand \Eprint [0]{\href }%
\providecommand \doibase [0]{http://dx.doi.org/}%
\providecommand \selectlanguage [0]{\@gobble}%
\providecommand \bibinfo  [0]{\@secondoftwo}%
\providecommand \bibfield  [0]{\@secondoftwo}%
\providecommand \translation [1]{[#1]}%
\providecommand \BibitemOpen [0]{}%
\providecommand \bibitemStop [0]{}%
\providecommand \bibitemNoStop [0]{.\EOS\space}%
\providecommand \EOS [0]{\spacefactor3000\relax}%
\providecommand \BibitemShut  [1]{\csname bibitem#1\endcsname}%
\let\auto@bib@innerbib\@empty
\bibitem [{\citenamefont {Majorana}(1937)}]{Majorana_1937}%
  \BibitemOpen
  \bibfield  {author} {\bibinfo {author} {\bibfnamefont {E.}~\bibnamefont
  {Majorana}},\ }\bibfield  {title} {\enquote {\bibinfo {title} {Teoria
  simmetrica dell'elettrone e del positrone},}\ }\href@noop {} {\bibfield
  {journal} {\bibinfo  {journal} {Nuovo Cimento}\ }\textbf {\bibinfo {volume}
  {14}},\ \bibinfo {pages} {171} (\bibinfo {year} {1937})}\BibitemShut
  {NoStop}%
\bibitem [{\citenamefont {Alicea}(2012)}]{Alicea_2012}%
  \BibitemOpen
  \bibfield  {author} {\bibinfo {author} {\bibfnamefont {J.}~\bibnamefont
  {Alicea}},\ }\bibfield  {title} {\enquote {\bibinfo {title} {New directions
  in the pursuit of {M}ajorana fermions in solid state systems},}\ }\href@noop
  {} {\bibfield  {journal} {\bibinfo  {journal} {Rep. Prog. Phys.}\ }\textbf
  {\bibinfo {volume} {75}},\ \bibinfo {pages} {076501} (\bibinfo {year}
  {2012})}\BibitemShut {NoStop}%
\bibitem [{\citenamefont {Leijnse}\ and\ \citenamefont
  {Flensberg}(2012)}]{Flensberg_2012}%
  \BibitemOpen
  \bibfield  {author} {\bibinfo {author} {\bibfnamefont {M.}~\bibnamefont
  {Leijnse}}\ and\ \bibinfo {author} {\bibfnamefont {K.}~\bibnamefont
  {Flensberg}},\ }\bibfield  {title} {\enquote {\bibinfo {title} {Introduction
  to topological superconductivity and {M}ajorana fermions},}\ }\href@noop {}
  {\bibfield  {journal} {\bibinfo  {journal} {Semicond. Sci. Technol.}\
  }\textbf {\bibinfo {volume} {27}},\ \bibinfo {pages} {124003} (\bibinfo
  {year} {2012})}\BibitemShut {NoStop}%
\bibitem [{\citenamefont {Sato}\ and\ \citenamefont
  {Fujimoto}(2016)}]{Sato_2016}%
  \BibitemOpen
  \bibfield  {author} {\bibinfo {author} {\bibfnamefont {M.}~\bibnamefont
  {Sato}}\ and\ \bibinfo {author} {\bibfnamefont {S.}~\bibnamefont
  {Fujimoto}},\ }\bibfield  {title} {\enquote {\bibinfo {title} {Majorana
  fermions and topology in superconductors},}\ }\href@noop {} {\bibfield
  {journal} {\bibinfo  {journal} {J. Phys. Soc. Japan}\ }\textbf {\bibinfo
  {volume} {85}},\ \bibinfo {pages} {072001} (\bibinfo {year}
  {2016})}\BibitemShut {NoStop}%
\bibitem [{\citenamefont {Aguado}(2017)}]{Aguado_2017}%
  \BibitemOpen
  \bibfield  {author} {\bibinfo {author} {\bibfnamefont {R.}~\bibnamefont
  {Aguado}},\ }\bibfield  {title} {\enquote {\bibinfo {title} {Majorana
  quasiparticles in condensed matter},}\ }\href@noop {} {\bibfield  {journal}
  {\bibinfo  {journal} {La Rivista del Nuovo Cimento}\ }\textbf {\bibinfo
  {volume} {40}},\ \bibinfo {pages} {523} (\bibinfo {year} {2017})}\BibitemShut
  {NoStop}%
\bibitem [{\citenamefont {Lutchyn}\ \emph {et~al.}(2018)\citenamefont
  {Lutchyn}, \citenamefont {Bakkers}, \citenamefont {Kouwenhoven},
  \citenamefont {Krogstrup}, \citenamefont {Marcus},\ and\ \citenamefont
  {Oreg}}]{Lutchyn_2018}%
  \BibitemOpen
  \bibfield  {author} {\bibinfo {author} {\bibfnamefont {R.~M.}\ \bibnamefont
  {Lutchyn}}, \bibinfo {author} {\bibfnamefont {E.~P. A.~M.}\ \bibnamefont
  {Bakkers}}, \bibinfo {author} {\bibfnamefont {L.~P.}\ \bibnamefont
  {Kouwenhoven}}, \bibinfo {author} {\bibfnamefont {P.}~\bibnamefont
  {Krogstrup}}, \bibinfo {author} {\bibfnamefont {C.~M.}\ \bibnamefont
  {Marcus}}, \ and\ \bibinfo {author} {\bibfnamefont {Y.}~\bibnamefont
  {Oreg}},\ }\bibfield  {title} {\enquote {\bibinfo {title} {Majorana zero
  modes in superconductor-semiconductor heterostructures},}\ }\href@noop {}
  {\bibfield  {journal} {\bibinfo  {journal} {Nat. Rev. Mater.}\ }\textbf
  {\bibinfo {volume} {3}},\ \bibinfo {pages} {52} (\bibinfo {year}
  {2018})}\BibitemShut {NoStop}%
\bibitem [{\citenamefont {\text{Yu.} Kitaev}(2001)}]{Kitaev_2001}%
  \BibitemOpen
  \bibfield  {author} {\bibinfo {author} {\bibfnamefont {A.}~\bibnamefont
  {\text{Yu.} Kitaev}},\ }\bibfield  {title} {\enquote {\bibinfo {title}
  {Unpaired {M}ajorana fermions in quantum wires},}\ }\href@noop {} {\bibfield
  {journal} {\bibinfo  {journal} {Phys.-Usp.}\ }\textbf {\bibinfo {volume}
  {44}},\ \bibinfo {pages} {131} (\bibinfo {year} {2001})}\BibitemShut
  {NoStop}%
\bibitem [{\citenamefont {Fu}\ and\ \citenamefont {Kane}(2008)}]{Fu_2008}%
  \BibitemOpen
  \bibfield  {author} {\bibinfo {author} {\bibfnamefont {L.}~\bibnamefont
  {Fu}}\ and\ \bibinfo {author} {\bibfnamefont {C.~L.}\ \bibnamefont {Kane}},\
  }\bibfield  {title} {\enquote {\bibinfo {title} {Superconducting proximity
  effect and {M}ajorana fermions at the surface of a topological insulator},}\
  }\href@noop {} {\bibfield  {journal} {\bibinfo  {journal} {Phys.\ Rev.\
  Lett.}\ }\textbf {\bibinfo {volume} {100}},\ \bibinfo {pages} {096407}
  (\bibinfo {year} {2008})}\BibitemShut {NoStop}%
\bibitem [{\citenamefont {Fu}\ and\ \citenamefont {Kane}(2009)}]{Fu_2009}%
  \BibitemOpen
  \bibfield  {author} {\bibinfo {author} {\bibfnamefont {L.}~\bibnamefont
  {Fu}}\ and\ \bibinfo {author} {\bibfnamefont {C.~L.}\ \bibnamefont {Kane}},\
  }\bibfield  {title} {\enquote {\bibinfo {title} {Josephson current and noise
  at a superconductor/quantum-spin-{H}all-insulator/superconductor junction},}\
  }\href@noop {} {\bibfield  {journal} {\bibinfo  {journal} {Phys.\ Rev.\ B}\
  }\textbf {\bibinfo {volume} {79}},\ \bibinfo {pages} {161408(R)} (\bibinfo
  {year} {2009})}\BibitemShut {NoStop}%
\bibitem [{\citenamefont {Lutchyn}\ \emph {et~al.}(2010)\citenamefont
  {Lutchyn}, \citenamefont {Sau},\ and\ \citenamefont {\text{Das
  Sarma}}}]{Lutchyn_2010}%
  \BibitemOpen
  \bibfield  {author} {\bibinfo {author} {\bibfnamefont {R.~M.}\ \bibnamefont
  {Lutchyn}}, \bibinfo {author} {\bibfnamefont {J.~D.}\ \bibnamefont {Sau}}, \
  and\ \bibinfo {author} {\bibfnamefont {S.}~\bibnamefont {\text{Das Sarma}}},\
  }\bibfield  {title} {\enquote {\bibinfo {title} {Majorana fermions and a
  topological phase transition in semiconductor-superconductor
  heterostructures},}\ }\href@noop {} {\bibfield  {journal} {\bibinfo
  {journal} {Phys.\ Rev.\ Lett.}\ }\textbf {\bibinfo {volume} {105}},\ \bibinfo
  {pages} {077001} (\bibinfo {year} {2010})}\BibitemShut {NoStop}%
\bibitem [{\citenamefont {Oreg}\ \emph {et~al.}(2010)\citenamefont {Oreg},
  \citenamefont {Refael},\ and\ \citenamefont {von Oppen}}]{Oreg_2010}%
  \BibitemOpen
  \bibfield  {author} {\bibinfo {author} {\bibfnamefont {Y.}~\bibnamefont
  {Oreg}}, \bibinfo {author} {\bibfnamefont {G.}~\bibnamefont {Refael}}, \ and\
  \bibinfo {author} {\bibfnamefont {F.}~\bibnamefont {von Oppen}},\ }\bibfield
  {title} {\enquote {\bibinfo {title} {Helical liquids and {M}ajorana bound
  states in quantum wires},}\ }\href@noop {} {\bibfield  {journal} {\bibinfo
  {journal} {Phys.\ Rev.\ Lett.}\ }\textbf {\bibinfo {volume} {105}},\ \bibinfo
  {pages} {177002} (\bibinfo {year} {2010})}\BibitemShut {NoStop}%
\bibitem [{\citenamefont {Mourik}\ \emph {et~al.}(2012)\citenamefont {Mourik},
  \citenamefont {Zuo}, \citenamefont {Frolov}, \citenamefont {Plissard},
  \citenamefont {Bakkers},\ and\ \citenamefont {Kouwenhoven}}]{Mourik_2012}%
  \BibitemOpen
  \bibfield  {author} {\bibinfo {author} {\bibfnamefont {V.}~\bibnamefont
  {Mourik}}, \bibinfo {author} {\bibfnamefont {K.}~\bibnamefont {Zuo}},
  \bibinfo {author} {\bibfnamefont {S.~M.}\ \bibnamefont {Frolov}}, \bibinfo
  {author} {\bibfnamefont {S.~R.}\ \bibnamefont {Plissard}}, \bibinfo {author}
  {\bibfnamefont {E.~P. A.~M.}\ \bibnamefont {Bakkers}}, \ and\ \bibinfo
  {author} {\bibfnamefont {L.~P.}\ \bibnamefont {Kouwenhoven}},\ }\bibfield
  {title} {\enquote {\bibinfo {title} {Signatures of {M}ajorana fermions in
  hybrid superconductor-semiconductor nanowire devices},}\ }\href@noop {}
  {\bibfield  {journal} {\bibinfo  {journal} {Science}\ }\textbf {\bibinfo
  {volume} {336}},\ \bibinfo {pages} {1003} (\bibinfo {year}
  {2012})}\BibitemShut {NoStop}%
\bibitem [{\citenamefont {Albrecht}\ \emph {et~al.}(2016)\citenamefont
  {Albrecht}, \citenamefont {Higginbotham}, \citenamefont {Madsen},
  \citenamefont {Kuemmeth}, \citenamefont {Jespersen}, \citenamefont
  {Nyg{\r{a}}rd}, \citenamefont {Krogstrup},\ and\ \citenamefont
  {Marcus}}]{Albrecht_2016}%
  \BibitemOpen
  \bibfield  {author} {\bibinfo {author} {\bibfnamefont {S.~M.}\ \bibnamefont
  {Albrecht}}, \bibinfo {author} {\bibfnamefont {A.~P.}\ \bibnamefont
  {Higginbotham}}, \bibinfo {author} {\bibfnamefont {M.}~\bibnamefont
  {Madsen}}, \bibinfo {author} {\bibfnamefont {F.}~\bibnamefont {Kuemmeth}},
  \bibinfo {author} {\bibfnamefont {T.~S.}\ \bibnamefont {Jespersen}}, \bibinfo
  {author} {\bibfnamefont {J.}~\bibnamefont {Nyg{\r{a}}rd}}, \bibinfo {author}
  {\bibfnamefont {P.}~\bibnamefont {Krogstrup}}, \ and\ \bibinfo {author}
  {\bibfnamefont {C.~M.}\ \bibnamefont {Marcus}},\ }\bibfield  {title}
  {\enquote {\bibinfo {title} {Exponential protection of zero modes in
  {M}ajorana islands},}\ }\href@noop {} {\bibfield  {journal} {\bibinfo
  {journal} {Nature}\ }\textbf {\bibinfo {volume} {531}},\ \bibinfo {pages}
  {206} (\bibinfo {year} {2016})}\BibitemShut {NoStop}%
\bibitem [{\citenamefont {Zhang}\ \emph {et~al.}(2018)\citenamefont {Zhang},
  \citenamefont {Liu}, \citenamefont {Gazibegovic}, \citenamefont {Xu},
  \citenamefont {Logan}, \citenamefont {Wang}, \citenamefont {van Loo},
  \citenamefont {Bommer}, \citenamefont {de~Moor}, \citenamefont {Car},
  \citenamefont {het Veld}, \citenamefont {van Veldhoven}, \citenamefont
  {Koelling}, \citenamefont {Verheijen}, \citenamefont {Pendharkar},
  \citenamefont {Pennachio}, \citenamefont {Shojaei}, \citenamefont {Lee},
  \citenamefont {Palmstr{\o}m}, \citenamefont {Bakkers}, \citenamefont
  {Sarma},\ and\ \citenamefont {Kouwenhoven}}]{Zhang_2018}%
  \BibitemOpen
  \bibfield  {author} {\bibinfo {author} {\bibfnamefont {H.}~\bibnamefont
  {Zhang}}, \bibinfo {author} {\bibfnamefont {C.-X.}\ \bibnamefont {Liu}},
  \bibinfo {author} {\bibfnamefont {S.}~\bibnamefont {Gazibegovic}}, \bibinfo
  {author} {\bibfnamefont {D.}~\bibnamefont {Xu}}, \bibinfo {author}
  {\bibfnamefont {J.~A.}\ \bibnamefont {Logan}}, \bibinfo {author}
  {\bibfnamefont {G.}~\bibnamefont {Wang}}, \bibinfo {author} {\bibfnamefont
  {N.}~\bibnamefont {van Loo}}, \bibinfo {author} {\bibfnamefont {J.~D.~S.}\
  \bibnamefont {Bommer}}, \bibinfo {author} {\bibfnamefont {M.~W.~A.}\
  \bibnamefont {de~Moor}}, \bibinfo {author} {\bibfnamefont {D.}~\bibnamefont
  {Car}}, \bibinfo {author} {\bibfnamefont {R.~L. M.~O.}\ \bibnamefont {het
  Veld}}, \bibinfo {author} {\bibfnamefont {P.~J.}\ \bibnamefont {van
  Veldhoven}}, \bibinfo {author} {\bibfnamefont {S.}~\bibnamefont {Koelling}},
  \bibinfo {author} {\bibfnamefont {M.~A.}\ \bibnamefont {Verheijen}}, \bibinfo
  {author} {\bibfnamefont {M.}~\bibnamefont {Pendharkar}}, \bibinfo {author}
  {\bibfnamefont {D.~J.}\ \bibnamefont {Pennachio}}, \bibinfo {author}
  {\bibfnamefont {B.}~\bibnamefont {Shojaei}}, \bibinfo {author} {\bibfnamefont
  {J.~S.}\ \bibnamefont {Lee}}, \bibinfo {author} {\bibfnamefont {C.~J.}\
  \bibnamefont {Palmstr{\o}m}}, \bibinfo {author} {\bibfnamefont {E.~P. A.~M.}\
  \bibnamefont {Bakkers}}, \bibinfo {author} {\bibfnamefont {S.~D.}\
  \bibnamefont {Sarma}}, \ and\ \bibinfo {author} {\bibfnamefont {L.~P.}\
  \bibnamefont {Kouwenhoven}},\ }\bibfield  {title} {\enquote {\bibinfo {title}
  {Quantized {M}ajorana conductance},}\ }\href@noop {} {\bibfield  {journal}
  {\bibinfo  {journal} {Nature}\ }\textbf {\bibinfo {volume} {556}},\ \bibinfo
  {pages} {74} (\bibinfo {year} {2018})}\BibitemShut {NoStop}%
\bibitem [{\citenamefont {Hartman}\ \emph {et~al.}(2018)\citenamefont
  {Hartman}, \citenamefont {Olsen}, \citenamefont {L{\"u}scher}, \citenamefont
  {Samani}, \citenamefont {Fallahi}, \citenamefont {Gardner}, \citenamefont
  {Manfra},\ and\ \citenamefont {Folk}}]{Hartman_2018}%
  \BibitemOpen
  \bibfield  {author} {\bibinfo {author} {\bibfnamefont {N.}~\bibnamefont
  {Hartman}}, \bibinfo {author} {\bibfnamefont {C.}~\bibnamefont {Olsen}},
  \bibinfo {author} {\bibfnamefont {S.}~\bibnamefont {L{\"u}scher}}, \bibinfo
  {author} {\bibfnamefont {M.}~\bibnamefont {Samani}}, \bibinfo {author}
  {\bibfnamefont {S.}~\bibnamefont {Fallahi}}, \bibinfo {author} {\bibfnamefont
  {G.~C.}\ \bibnamefont {Gardner}}, \bibinfo {author} {\bibfnamefont
  {M.}~\bibnamefont {Manfra}}, \ and\ \bibinfo {author} {\bibfnamefont
  {J.}~\bibnamefont {Folk}},\ }\bibfield  {title} {\enquote {\bibinfo {title}
  {Direct entropy measurement in a mesoscopic quantum system},}\ }\href@noop {}
  {\bibfield  {journal} {\bibinfo  {journal} {Nature Physics}\ }\textbf
  {\bibinfo {volume} {14}},\ \bibinfo {pages} {1083} (\bibinfo {year}
  {2018})}\BibitemShut {NoStop}%
\bibitem [{\citenamefont {Kleeorin}\ \emph {et~al.}(2019)\citenamefont
  {Kleeorin}, \citenamefont {Thierschmann}, \citenamefont {Buhmann},
  \citenamefont {Georges}, \citenamefont {Molenkamp},\ and\ \citenamefont
  {Meir}}]{Kleeorin_2019}%
  \BibitemOpen
  \bibfield  {author} {\bibinfo {author} {\bibfnamefont {Y.}~\bibnamefont
  {Kleeorin}}, \bibinfo {author} {\bibfnamefont {H.}~\bibnamefont
  {Thierschmann}}, \bibinfo {author} {\bibfnamefont {H.}~\bibnamefont
  {Buhmann}}, \bibinfo {author} {\bibfnamefont {A.}~\bibnamefont {Georges}},
  \bibinfo {author} {\bibfnamefont {L.~W.}\ \bibnamefont {Molenkamp}}, \ and\
  \bibinfo {author} {\bibfnamefont {Y.}~\bibnamefont {Meir}},\ }\bibfield
  {title} {\enquote {\bibinfo {title} {Measuring the entropy of a mesoscopic
  system via thermoelectric transport},}\ }\href@noop {} {\bibfield  {journal}
  {\bibinfo  {journal} {arXiv:1904.08948}\ } (\bibinfo {year}
  {2019})}\BibitemShut {NoStop}%
\bibitem [{\citenamefont {Sela}\ \emph {et~al.}(2019)\citenamefont {Sela},
  \citenamefont {Oreg}, \citenamefont {Plugge}, \citenamefont {Hartman},
  \citenamefont {L{\"u}scher},\ and\ \citenamefont {Folk}}]{Sela_2019}%
  \BibitemOpen
  \bibfield  {author} {\bibinfo {author} {\bibfnamefont {E.}~\bibnamefont
  {Sela}}, \bibinfo {author} {\bibfnamefont {Y.}~\bibnamefont {Oreg}}, \bibinfo
  {author} {\bibfnamefont {S.}~\bibnamefont {Plugge}}, \bibinfo {author}
  {\bibfnamefont {N.}~\bibnamefont {Hartman}}, \bibinfo {author} {\bibfnamefont
  {S.}~\bibnamefont {L{\"u}scher}}, \ and\ \bibinfo {author} {\bibfnamefont
  {J.}~\bibnamefont {Folk}},\ }\bibfield  {title} {\enquote {\bibinfo {title}
  {Detecting the universal fractional entropy of {M}ajorana zero modes},}\
  }\href@noop {} {\bibfield  {journal} {\bibinfo  {journal} {Phys.\ Rev.\
  Lett.}\ }\textbf {\bibinfo {volume} {123}},\ \bibinfo {pages} {147702}
  (\bibinfo {year} {2019})}\BibitemShut {NoStop}%
\bibitem [{\citenamefont {Smirnov}(2015)}]{Smirnov_2015}%
  \BibitemOpen
  \bibfield  {author} {\bibinfo {author} {\bibfnamefont {S.}~\bibnamefont
  {Smirnov}},\ }\bibfield  {title} {\enquote {\bibinfo {title} {Majorana
  tunneling entropy},}\ }\href@noop {} {\bibfield  {journal} {\bibinfo
  {journal} {Phys.\ Rev.\ B}\ }\textbf {\bibinfo {volume} {92}},\ \bibinfo
  {pages} {195312} (\bibinfo {year} {2015})}\BibitemShut {NoStop}%
\bibitem [{\citenamefont {\text{Yu.} Kitaev}(2003)}]{Kitaev_2003}%
  \BibitemOpen
  \bibfield  {author} {\bibinfo {author} {\bibfnamefont {A.}~\bibnamefont
  {\text{Yu.} Kitaev}},\ }\bibfield  {title} {\enquote {\bibinfo {title}
  {Fault-tolerant quantum computation by anyons},}\ }\href@noop {} {\bibfield
  {journal} {\bibinfo  {journal} {Ann. Phys.}\ }\textbf {\bibinfo {volume}
  {303}},\ \bibinfo {pages} {2} (\bibinfo {year} {2003})}\BibitemShut {NoStop}%
\bibitem [{\citenamefont {Nayak}\ \emph {et~al.}(2008)\citenamefont {Nayak},
  \citenamefont {Simon}, \citenamefont {Stern}, \citenamefont {Freedman},\ and\
  \citenamefont {\text{Das Sarma}}}]{Nayak_2008}%
  \BibitemOpen
  \bibfield  {author} {\bibinfo {author} {\bibfnamefont {C.}~\bibnamefont
  {Nayak}}, \bibinfo {author} {\bibfnamefont {S.~H.}\ \bibnamefont {Simon}},
  \bibinfo {author} {\bibfnamefont {A.}~\bibnamefont {Stern}}, \bibinfo
  {author} {\bibfnamefont {M.}~\bibnamefont {Freedman}}, \ and\ \bibinfo
  {author} {\bibfnamefont {S.}~\bibnamefont {\text{Das Sarma}}},\ }\bibfield
  {title} {\enquote {\bibinfo {title} {Non-abelian anyons and topological
  quantum computation},}\ }\href@noop {} {\bibfield  {journal} {\bibinfo
  {journal} {Rev.\ Mod.\ Phys.}\ }\textbf {\bibinfo {volume} {80}},\ \bibinfo
  {pages} {1083} (\bibinfo {year} {2008})}\BibitemShut {NoStop}%
\bibitem [{\citenamefont {Pedrocchi}\ and\ \citenamefont
  {DiVincenzo}(2015)}]{Pedrocchi_2015}%
  \BibitemOpen
  \bibfield  {author} {\bibinfo {author} {\bibfnamefont {F.~L.}\ \bibnamefont
  {Pedrocchi}}\ and\ \bibinfo {author} {\bibfnamefont {D.~P.}\ \bibnamefont
  {DiVincenzo}},\ }\bibfield  {title} {\enquote {\bibinfo {title} {Majorana
  braiding with thermal noise},}\ }\href@noop {} {\bibfield  {journal}
  {\bibinfo  {journal} {Phys.\ Rev.\ Lett.}\ }\textbf {\bibinfo {volume}
  {115}},\ \bibinfo {pages} {120402} (\bibinfo {year} {2015})}\BibitemShut
  {NoStop}%
\bibitem [{\citenamefont {Liu}\ \emph {et~al.}(2015{\natexlab{a}})\citenamefont
  {Liu}, \citenamefont {Cheng},\ and\ \citenamefont {Lutchyn}}]{Liu_2015}%
  \BibitemOpen
  \bibfield  {author} {\bibinfo {author} {\bibfnamefont {D.~E.}\ \bibnamefont
  {Liu}}, \bibinfo {author} {\bibfnamefont {M.}~\bibnamefont {Cheng}}, \ and\
  \bibinfo {author} {\bibfnamefont {R.~M.}\ \bibnamefont {Lutchyn}},\
  }\bibfield  {title} {\enquote {\bibinfo {title} {Probing {M}ajorana physics
  in quantum-dot shot-noise experiments},}\ }\href@noop {} {\bibfield
  {journal} {\bibinfo  {journal} {Phys.\ Rev.\ B}\ }\textbf {\bibinfo {volume}
  {91}},\ \bibinfo {pages} {081405(R)} (\bibinfo {year}
  {2015}{\natexlab{a}})}\BibitemShut {NoStop}%
\bibitem [{\citenamefont {Liu}\ \emph {et~al.}(2015{\natexlab{b}})\citenamefont
  {Liu}, \citenamefont {Levchenko},\ and\ \citenamefont {Lutchyn}}]{Liu_2015a}%
  \BibitemOpen
  \bibfield  {author} {\bibinfo {author} {\bibfnamefont {D.~E.}\ \bibnamefont
  {Liu}}, \bibinfo {author} {\bibfnamefont {A.}~\bibnamefont {Levchenko}}, \
  and\ \bibinfo {author} {\bibfnamefont {R.~M.}\ \bibnamefont {Lutchyn}},\
  }\bibfield  {title} {\enquote {\bibinfo {title} {Majorana zero modes choose
  {E}uler numbers as revealed by full counting statistics},}\ }\href@noop {}
  {\bibfield  {journal} {\bibinfo  {journal} {Phys.\ Rev.\ B}\ }\textbf
  {\bibinfo {volume} {92}},\ \bibinfo {pages} {205422} (\bibinfo {year}
  {2015}{\natexlab{b}})}\BibitemShut {NoStop}%
\bibitem [{\citenamefont {Beenakker}(2015)}]{Beenakker_2015}%
  \BibitemOpen
  \bibfield  {author} {\bibinfo {author} {\bibfnamefont {C.~W.~J.}\
  \bibnamefont {Beenakker}},\ }\bibfield  {title} {\enquote {\bibinfo {title}
  {Random-matrix theory of {M}ajorana fermions and topological
  superconductors},}\ }\href@noop {} {\bibfield  {journal} {\bibinfo  {journal}
  {Rev.\ Mod.\ Phys.}\ }\textbf {\bibinfo {volume} {87}},\ \bibinfo {pages}
  {1037} (\bibinfo {year} {2015})}\BibitemShut {NoStop}%
\bibitem [{\citenamefont {Haim}\ \emph {et~al.}(2015)\citenamefont {Haim},
  \citenamefont {Berg}, \citenamefont {von Oppen},\ and\ \citenamefont
  {Oreg}}]{Haim_2015}%
  \BibitemOpen
  \bibfield  {author} {\bibinfo {author} {\bibfnamefont {A.}~\bibnamefont
  {Haim}}, \bibinfo {author} {\bibfnamefont {E.}~\bibnamefont {Berg}}, \bibinfo
  {author} {\bibfnamefont {F.}~\bibnamefont {von Oppen}}, \ and\ \bibinfo
  {author} {\bibfnamefont {Y.}~\bibnamefont {Oreg}},\ }\bibfield  {title}
  {\enquote {\bibinfo {title} {Current correlations in a {M}ajorana beam
  splitter},}\ }\href@noop {} {\bibfield  {journal} {\bibinfo  {journal}
  {Phys.\ Rev.\ B}\ }\textbf {\bibinfo {volume} {92}},\ \bibinfo {pages}
  {245112} (\bibinfo {year} {2015})}\BibitemShut {NoStop}%
\bibitem [{\citenamefont {Smirnov}(2017)}]{Smirnov_2017}%
  \BibitemOpen
  \bibfield  {author} {\bibinfo {author} {\bibfnamefont {S.}~\bibnamefont
  {Smirnov}},\ }\bibfield  {title} {\enquote {\bibinfo {title} {Non-equilibrium
  {M}ajorana fluctuations},}\ }\href@noop {} {\bibfield  {journal} {\bibinfo
  {journal} {New J. Phys.}\ }\textbf {\bibinfo {volume} {19}},\ \bibinfo
  {pages} {063020} (\bibinfo {year} {2017})}\BibitemShut {NoStop}%
\bibitem [{\citenamefont {Valentini}\ \emph {et~al.}(2016)\citenamefont
  {Valentini}, \citenamefont {Governale}, \citenamefont {Fazio},\ and\
  \citenamefont {Taddei}}]{Valentini_2016}%
  \BibitemOpen
  \bibfield  {author} {\bibinfo {author} {\bibfnamefont {S.}~\bibnamefont
  {Valentini}}, \bibinfo {author} {\bibfnamefont {M.}~\bibnamefont
  {Governale}}, \bibinfo {author} {\bibfnamefont {R.}~\bibnamefont {Fazio}}, \
  and\ \bibinfo {author} {\bibfnamefont {F.}~\bibnamefont {Taddei}},\
  }\bibfield  {title} {\enquote {\bibinfo {title} {Finite-frequency noise in a
  topological superconducting wire},}\ }\href@noop {} {\bibfield  {journal}
  {\bibinfo  {journal} {Physica E}\ }\textbf {\bibinfo {volume} {75}},\
  \bibinfo {pages} {15} (\bibinfo {year} {2016})}\BibitemShut {NoStop}%
\bibitem [{\citenamefont {Bathellier}\ \emph {et~al.}(2019)\citenamefont
  {Bathellier}, \citenamefont {Raymond}, \citenamefont {Jonckheere},
  \citenamefont {Rech}, \citenamefont {Zazunov},\ and\ \citenamefont
  {Martin}}]{Bathellier_2019}%
  \BibitemOpen
  \bibfield  {author} {\bibinfo {author} {\bibfnamefont {D.}~\bibnamefont
  {Bathellier}}, \bibinfo {author} {\bibfnamefont {L.}~\bibnamefont {Raymond}},
  \bibinfo {author} {\bibfnamefont {T.}~\bibnamefont {Jonckheere}}, \bibinfo
  {author} {\bibfnamefont {J.}~\bibnamefont {Rech}}, \bibinfo {author}
  {\bibfnamefont {A.}~\bibnamefont {Zazunov}}, \ and\ \bibinfo {author}
  {\bibfnamefont {T.}~\bibnamefont {Martin}},\ }\bibfield  {title} {\enquote
  {\bibinfo {title} {Finite frequency noise in a normal metal - topological
  superconductor junction},}\ }\href@noop {} {\bibfield  {journal} {\bibinfo
  {journal} {Phys.\ Rev.\ B}\ }\textbf {\bibinfo {volume} {99}},\ \bibinfo
  {pages} {104502} (\bibinfo {year} {2019})}\BibitemShut {NoStop}%
\bibitem [{\citenamefont {Smirnov}(2019)}]{Smirnov_2019}%
  \BibitemOpen
  \bibfield  {author} {\bibinfo {author} {\bibfnamefont {S.}~\bibnamefont
  {Smirnov}},\ }\bibfield  {title} {\enquote {\bibinfo {title} {Majorana
  finite-frequency nonequilibrium quantum noise},}\ }\href@noop {} {\bibfield
  {journal} {\bibinfo  {journal} {Phys.\ Rev.\ B}\ }\textbf {\bibinfo {volume}
  {99}},\ \bibinfo {pages} {165427} (\bibinfo {year} {2019})}\BibitemShut
  {NoStop}%
\bibitem [{\citenamefont {Smirnov}(2018)}]{Smirnov_2018}%
  \BibitemOpen
  \bibfield  {author} {\bibinfo {author} {\bibfnamefont {S.}~\bibnamefont
  {Smirnov}},\ }\bibfield  {title} {\enquote {\bibinfo {title} {Universal
  {M}ajorana thermoelectric noise},}\ }\href@noop {} {\bibfield  {journal}
  {\bibinfo  {journal} {Phys.\ Rev.\ B}\ }\textbf {\bibinfo {volume} {97}},\
  \bibinfo {pages} {165434} (\bibinfo {year} {2018})}\BibitemShut {NoStop}%
\bibitem [{\citenamefont {Clerk}\ \emph {et~al.}(2010)\citenamefont {Clerk},
  \citenamefont {Devoret}, \citenamefont {Girvin}, \citenamefont {Marquardt},\
  and\ \citenamefont {Schoelkopf}}]{Clerk_2010}%
  \BibitemOpen
  \bibfield  {author} {\bibinfo {author} {\bibfnamefont {A.~A.}\ \bibnamefont
  {Clerk}}, \bibinfo {author} {\bibfnamefont {M.~H.}\ \bibnamefont {Devoret}},
  \bibinfo {author} {\bibfnamefont {S.~M.}\ \bibnamefont {Girvin}}, \bibinfo
  {author} {\bibfnamefont {F.}~\bibnamefont {Marquardt}}, \ and\ \bibinfo
  {author} {\bibfnamefont {R.~J.}\ \bibnamefont {Schoelkopf}},\ }\bibfield
  {title} {\enquote {\bibinfo {title} {Introduction to quantum noise,
  measurement, and amplification},}\ }\href@noop {} {\bibfield  {journal}
  {\bibinfo  {journal} {Rev.\ Mod.\ Phys.}\ }\textbf {\bibinfo {volume} {82}},\
  \bibinfo {pages} {1155} (\bibinfo {year} {2010})}\BibitemShut {NoStop}%
\bibitem [{\citenamefont {Altland}\ and\ \citenamefont
  {Simons}(2010)}]{Altland_2010}%
  \BibitemOpen
  \bibfield  {author} {\bibinfo {author} {\bibfnamefont {A.}~\bibnamefont
  {Altland}}\ and\ \bibinfo {author} {\bibfnamefont {B.}~\bibnamefont
  {Simons}},\ }\href@noop {} {\emph {\bibinfo {title} {Condensed Matter Field
  Theory}}},\ \bibinfo {edition} {2nd}\ ed.\ (\bibinfo  {publisher} {Cambridge
  University Press, Cambridge},\ \bibinfo {year} {2010})\BibitemShut {NoStop}%
\bibitem [{\citenamefont {Lesovik}\ and\ \citenamefont
  {Loosen}(1997)}]{Lesovik_1997}%
  \BibitemOpen
  \bibfield  {author} {\bibinfo {author} {\bibfnamefont {G.~B.}\ \bibnamefont
  {Lesovik}}\ and\ \bibinfo {author} {\bibfnamefont {R.}~\bibnamefont
  {Loosen}},\ }\bibfield  {title} {\enquote {\bibinfo {title} {On the detection
  of finite-frequency current fluctuations},}\ }\href@noop {} {\bibfield
  {journal} {\bibinfo  {journal} {JETP Lett.}\ }\textbf {\bibinfo {volume}
  {65}},\ \bibinfo {pages} {295} (\bibinfo {year} {1997})}\BibitemShut
  {NoStop}%
\bibitem [{\citenamefont {Gavish}\ \emph {et~al.}(2000)\citenamefont {Gavish},
  \citenamefont {Levinson},\ and\ \citenamefont {Imry}}]{Gavish_2000}%
  \BibitemOpen
  \bibfield  {author} {\bibinfo {author} {\bibfnamefont {U.}~\bibnamefont
  {Gavish}}, \bibinfo {author} {\bibfnamefont {Y.}~\bibnamefont {Levinson}}, \
  and\ \bibinfo {author} {\bibfnamefont {Y.}~\bibnamefont {Imry}},\ }\bibfield
  {title} {\enquote {\bibinfo {title} {Detection of quantum noise},}\
  }\href@noop {} {\bibfield  {journal} {\bibinfo  {journal} {Phys.\ Rev.\ B}\
  }\textbf {\bibinfo {volume} {62}},\ \bibinfo {pages} {R10637(R)} (\bibinfo
  {year} {2000})}\BibitemShut {NoStop}%
\bibitem [{\citenamefont {Moore}\ \emph {et~al.}(2018)\citenamefont {Moore},
  \citenamefont {Stanescu},\ and\ \citenamefont {Tewari}}]{Moore_2018}%
  \BibitemOpen
  \bibfield  {author} {\bibinfo {author} {\bibfnamefont {C.}~\bibnamefont
  {Moore}}, \bibinfo {author} {\bibfnamefont {T.~D.}\ \bibnamefont {Stanescu}},
  \ and\ \bibinfo {author} {\bibfnamefont {S.}~\bibnamefont {Tewari}},\
  }\bibfield  {title} {\enquote {\bibinfo {title} {Two-terminal charge
  tunneling: {D}isentangling {M}ajorana zero modes from partially separated
  {A}ndreev bound states in semiconductor-superconductor heterostructures},}\
  }\href@noop {} {\bibfield  {journal} {\bibinfo  {journal} {Phys.\ Rev.\ B}\
  }\textbf {\bibinfo {volume} {97}},\ \bibinfo {pages} {165302} (\bibinfo
  {year} {2018})}\BibitemShut {NoStop}%
\bibitem [{\citenamefont {Hell}\ \emph {et~al.}(2018)\citenamefont {Hell},
  \citenamefont {Flensberg},\ and\ \citenamefont {Leijnse}}]{Hell_2018}%
  \BibitemOpen
  \bibfield  {author} {\bibinfo {author} {\bibfnamefont {M.}~\bibnamefont
  {Hell}}, \bibinfo {author} {\bibfnamefont {K.}~\bibnamefont {Flensberg}}, \
  and\ \bibinfo {author} {\bibfnamefont {M.}~\bibnamefont {Leijnse}},\
  }\bibfield  {title} {\enquote {\bibinfo {title} {Distinguishing {M}ajorana
  bound states from localized {A}ndreev bound states by interferometry},}\
  }\href@noop {} {\bibfield  {journal} {\bibinfo  {journal} {Phys.\ Rev.\ B}\
  }\textbf {\bibinfo {volume} {97}},\ \bibinfo {pages} {161401(R)} (\bibinfo
  {year} {2018})}\BibitemShut {NoStop}%
\bibitem [{\citenamefont {Clarke}(2017)}]{Clarke_2017}%
  \BibitemOpen
  \bibfield  {author} {\bibinfo {author} {\bibfnamefont {D.~J.}\ \bibnamefont
  {Clarke}},\ }\bibfield  {title} {\enquote {\bibinfo {title} {Experimentally
  accessible topological quality factor for wires with zero energy modes},}\
  }\href@noop {} {\bibfield  {journal} {\bibinfo  {journal} {Phys.\ Rev.\ B}\
  }\textbf {\bibinfo {volume} {96}},\ \bibinfo {pages} {201109(R)} (\bibinfo
  {year} {2017})}\BibitemShut {NoStop}%
\bibitem [{\citenamefont {Deng}\ \emph {et~al.}(2018)\citenamefont {Deng},
  \citenamefont {Vaitiek{\.{e}}nas}, \citenamefont {Prada}, \citenamefont
  {San-Jose}, \citenamefont {Nyg{\r{a}}rd}, \citenamefont {Krogstrup},
  \citenamefont {Aguado},\ and\ \citenamefont {Marcus}}]{Deng_2018}%
  \BibitemOpen
  \bibfield  {author} {\bibinfo {author} {\bibfnamefont {M.-T.}\ \bibnamefont
  {Deng}}, \bibinfo {author} {\bibfnamefont {S.}~\bibnamefont
  {Vaitiek{\.{e}}nas}}, \bibinfo {author} {\bibfnamefont {E.}~\bibnamefont
  {Prada}}, \bibinfo {author} {\bibfnamefont {P.}~\bibnamefont {San-Jose}},
  \bibinfo {author} {\bibfnamefont {J.}~\bibnamefont {Nyg{\r{a}}rd}}, \bibinfo
  {author} {\bibfnamefont {P.}~\bibnamefont {Krogstrup}}, \bibinfo {author}
  {\bibfnamefont {R.}~\bibnamefont {Aguado}}, \ and\ \bibinfo {author}
  {\bibfnamefont {C.~M.}\ \bibnamefont {Marcus}},\ }\bibfield  {title}
  {\enquote {\bibinfo {title} {Nonlocality of {M}ajorana modes in hybrid
  nanowires},}\ }\href@noop {} {\bibfield  {journal} {\bibinfo  {journal}
  {Phys.\ Rev.\ B}\ }\textbf {\bibinfo {volume} {98}},\ \bibinfo {pages}
  {085125} (\bibinfo {year} {2018})}\BibitemShut {NoStop}%
\bibitem [{\citenamefont {Averin}(2000)}]{Averin_2000}%
  \BibitemOpen
  \bibfield  {author} {\bibinfo {author} {\bibfnamefont {D.~V.}\ \bibnamefont
  {Averin}},\ }\bibfield  {title} {\enquote {\bibinfo {title} {Quantum
  computing and quantum measurement with mesoscopic {J}osephson junctions},}\
  }\href@noop {} {\bibfield  {journal} {\bibinfo  {journal} {Fortschr. Phys.}\
  }\textbf {\bibinfo {volume} {48}},\ \bibinfo {pages} {1055} (\bibinfo {year}
  {2000})}\BibitemShut {NoStop}%
\bibitem [{\citenamefont {Clerk}\ and\ \citenamefont
  {Stone}(2004)}]{Clerk_2004}%
  \BibitemOpen
  \bibfield  {author} {\bibinfo {author} {\bibfnamefont {A.~A.}\ \bibnamefont
  {Clerk}}\ and\ \bibinfo {author} {\bibfnamefont {A.~D.}\ \bibnamefont
  {Stone}},\ }\bibfield  {title} {\enquote {\bibinfo {title} {Noise and
  measurement efficiency of a partially coherent mesoscopic detector},}\
  }\href@noop {} {\bibfield  {journal} {\bibinfo  {journal} {Phys.\ Rev.\ B}\
  }\textbf {\bibinfo {volume} {69}},\ \bibinfo {pages} {245303} (\bibinfo
  {year} {2004})}\BibitemShut {NoStop}%
\bibitem [{\citenamefont {Mozyrsky}\ \emph {et~al.}(2004)\citenamefont
  {Mozyrsky}, \citenamefont {Martin},\ and\ \citenamefont
  {Hastings}}]{Mozyrsky_2004}%
  \BibitemOpen
  \bibfield  {author} {\bibinfo {author} {\bibfnamefont {D.}~\bibnamefont
  {Mozyrsky}}, \bibinfo {author} {\bibfnamefont {I.}~\bibnamefont {Martin}}, \
  and\ \bibinfo {author} {\bibfnamefont {M.~B.}\ \bibnamefont {Hastings}},\
  }\bibfield  {title} {\enquote {\bibinfo {title} {Quantum-limited sensitivity
  of single-electron-transistor-based displacement detectors},}\ }\href@noop {}
  {\bibfield  {journal} {\bibinfo  {journal} {Phys.\ Rev.\ Lett.}\ }\textbf
  {\bibinfo {volume} {92}},\ \bibinfo {pages} {018303} (\bibinfo {year}
  {2004})}\BibitemShut {NoStop}%
\end{thebibliography}
\end{document}